\documentclass[]{iopart}

\usepackage{amsfonts}
\usepackage{amssymb}
\usepackage{graphicx}
\usepackage{moreverb}
\usepackage{url}
\usepackage{epsf}
\usepackage{color}
\usepackage{multirow}
\usepackage{subfigure}
\usepackage{cite}
\usepackage{overpic}

\makeatletter
\let\gplgaddtomacro\g@addto@macro
\makeatother
\begin{document}

\title{A first principles study of As doping at a disordered S\lowercase{i}--S\lowercase{i}O$_2$ interface}
\author{Fabiano~Corsetti$^{1,2}$ and Arash~A~Mostofi$^2$}
\address{$^1$ CIC nanoGUNE, 20018 Donostia-San Sebasti\'{a}n, Spain}
\address{$^2$ Departments of Materials and Physics, and the Thomas Young Centre for Theory and Simulation of Materials, Imperial College London, London SW7 2AZ, United Kingdom}
\eads{f.corsetti@nanogune.eu}

\begin{abstract}
Understanding the interaction between dopants and semiconductor-oxide interfaces is an increasingly important concern in the drive to further miniaturize modern transistors. To this end, using a combination of first-principles density-functional theory and a continuous random network Monte Carlo method, we investigate electrically active arsenic donors at the interface between silicon and its oxide. Using a realistic model of the disordered interface, we find that a small percentage (on the order of $\sim$10\%) of the atomic sites in the first few monolayers on the silicon side of the interface are energetically favourable for segregation, and that this is controlled by the local bonding and local strain of the defect centre. We also find that there is a long-range quantum confinement effect due to the interface, which results in an energy barrier for dopant segregation, but that this barrier is small in comparison to the effect of the local environment. Finally, we consider the extent to which the energetics of segregation can be controlled by the application of strain to the interface.
\end{abstract}

\maketitle

\section{Introduction}
\label{sec:intro}

The Si--SiO$_2$ interface is a common feature in modern silicon-based complementary metal-oxide-semiconductor (CMOS) technology for the fabrication of integrated circuits. In the drive to scale devices to smaller sizes, channel lengths have been reduced to a few tens of nanometres~\cite{MOSFET-evol}, with the properties of the device being determined by fewer than 100 dopant atoms~\cite{arsenic-ionization}. Understanding the effects of the interface on dopants is therefore an increasingly important concern.

Several experimental studies have reported the uphill thermal diffusion and segregation of arsenic atoms to the Si--SiO$_2$ interface region during the high-temperature anneal following implantation~\cite{Sb-As-uphill,As-uphill,As-uphill2,As-B-uphill,As-B-uphill2,As-uphill3,As-uphill4,As-uphill5,As-uphill6,As-uphill7}. Up to a monolayer of dopant atoms can be collected in this region~\cite{Sb-As-uphill,uphill-model}. Studies using X-ray photoelectron spectroscopy (XPS)~\cite{As-uphill}, medium ion scattering (MEIS)~\cite{As-B-uphill} and scanning transmission electron microscopy (STEM)~\cite{As-uphill3} have shown the dopants to be on the silicon side of the interface, despite initial suggestions to the contrary~\cite{pileup-oxide,pileup-oxide2}. 

While the majority of segregated dopants are deactivated~\cite{Sb-As-uphill,pileup-neutral}, giving rise to significant dose loss in the device and affecting the threshold voltage by up to 50\%~\cite{pileup-threshold}, 10--20\% of them are estimated to be active~\cite{pileup-active}. In the context of device miniaturization, understanding the properties of active dopants close to the interface becomes crucial~\cite{arsenic-ionization, nanowire-ionization}.

In this paper, we use first-principles density-functional theory (DFT) simulations to study arsenic impurities at Si--SiO$_2$ interfaces. In contrast to previous first-principles calculations~\cite{seg-comp,seg-comp2,seg-comp3,seg-comp4,seg-comp5,seg-comp6} that focus on defect configurations within a monolayer of the interface that passivate the donor and favour segregation, our emphasis is on understanding electrically active arsenic impurities in bulk-like, four-fold coordinated configurations at and near the interface.

The starting point of developing a good atomistic model of the Si--SiO$_2$ interface is itself a challenging problem. Previous studies have used either hand-built models~\cite{Si-SiO2-car,Si-SiO2-car2,Si-SiO2-car3,Si-SiO2-hand,Si-SiO2-hand2}, classical~\cite{Si-SiO2-md} or {\em ab initio}~\cite{Si-SiO2-car4} molecular dynamics (MD), or Monte Carlo (MC) methods~\cite{Si-SiO2-crn,Si-SiO2-crn2,Si-SiO2-crn3}. We use a large supercell of 472 atoms and consider both an idealized ordered interface of crystalline Si with one of the crystalline phases of SiO$_2$, and a more realistic disordered interface of crystalline Si with amorphous SiO$_2$. The former is generated by lattice matching the two crystals and introducing bridge atoms at the interface to ensure full coordination, while the latter is generated using a multiscale approach, based on a continuous random network Monte Carlo (CRN-MC) model parametrized by DFT.

The large interfacial area in our disordered interface model enables us to place the As impurity atom at many inequivalent sites close to the interface, in order to build up a representative picture of the segregation of active dopants in the system. Our main findings are that segregation is favoured to a small number of fully Si-coordinated sites, and that the local atomic environment of the defect site has a substantial effect on the segregation energy. While there is also a long-range quantum confinement effect due to the proximity of the dopant to the interface, we find that this is a small contribution to the segregation energy. Lattice relaxation is found to further enhance impurity segregation without causing donor passivation, although this effect is not sufficiently large to overcome the energy penalty associated with impurities being placed at partially or fully O-coordinated sites. Finally, we find that it may be possible to tune the segregation of dopants at the interface by applying macroscopic strain to the system.

The rest of the paper is organized as follows: in Sec.~\ref{sec:methods}, we first describe the computational techniques that we employ for generating the crystalline (Sec.~\ref{subsubsec:methods-interface-crystal}) and disordered (Sec.~\ref{subsubsec:methods-interface-disorder}) interface supercells, and then give the technical details of our DFT simulations of the As substitutional defect close to and at the interface (Sec.~\ref{subsec:methods-As}). In Sec.~\ref{sec:results}, we present our results; we first discuss our simulations of the dopant in the crystalline system (Sec.~\ref{subsec:results-crystal}) and in the disordered one (Sec.~\ref{subsec:results-disorder}), and then we discuss the effect of lattice relaxation (Sec.~\ref{subsec:results-rel}) and macroscopic strain (Sec.~\ref{subsec:results-strain}) in both systems. Finally, in Sec.~\ref{sec:outro}, we give a summary of our main conclusions.

\section{Methods}
\label{sec:methods}

\subsection{Si--SiO$_2$ interface generation}
\label{subsec:methods-interface}

\subsubsection{The crystalline interface}
\label{subsubsec:methods-interface-crystal}

Our model of a crystalline interface between Si and SiO$_2$ was constructed using the method of Pasquarello \etal~\cite{Si-SiO2-car,Si-SiO2-car2,Si-SiO2-car3}. The supercell was obtained by attaching cells of $\alpha$-cristobalite~\cite{Si-SiO2-crn3} (with an 8\% tensile strain for lattice matching) to a $2 \times 2$ Si(001) surface. The bond density mismatch at the interface was corrected with the addition of O bridges between atoms in the surface layer of Si (two per Si cubic cell), thus saturating the remaining dangling bonds (see top panel of Fig.~\ref{fig:seg-ideal}). The resulting tetragonal supercell, with 33 monolayers of Si and 8 monolayers of SiO$_2$, was then fully relaxed (ionic positions and lattice vectors) with DFT, starting from a preliminary relaxation obtained on a smaller supercell with only 9 monolayers of Si. For this we used the {\sc castep}~\cite{castep} code, with norm-conserving pseudopotentials~\cite{norm-pseudo} in Rappe-Rabe-Kaxiras-Joannopoulos (RRKJ) form~\cite{rrkj-pseudo}, a plane-wave basis with a kinetic energy cutoff of 800~eV, $\Gamma$-point sampling of the Brillouin zone, and the Ceperley-Alder~\cite{qmc1} local-density approximation (LDA) for describing exchange and correlation. Our convergence tolerance parameters were $10^{-2}$~eV/\AA\ for the maximum ionic force, $10^{-3}$~\AA\ for the maximum ionic displacement between iterations, $10^{-7}$~eV for the change in the total energy/ion between iterations, and $10^{-2}$~GPa for the maximum component of the stress tensor.

The relaxation produced no noticeable change in the atomic configuration, and only a small decrease of the size of the supercell perpendicular to the interface plane, resulting in a negligible increase in density for SiO$_2$. The final widths of the Si and SiO$_2$ layers were $\sim$42.5~\AA\ and $\sim$29.6~\AA, respectively~\footnote{The $\alpha$-cristobalite interface includes only the $\mathrm{Si}^{2+}$ oxidation state, in contrast with experiment. We have also constructed a similar ordered model using $\alpha$-tridymite~\cite{Si-SiO2-car2}, resulting in all the suboxide states being present at the interface; however, it is not possible in this case to achieve an equivalent bonding topology at both interfaces present in the supercell due to the periodic boundary conditions, resulting in a large charge transfer of almost one electron per Si cubic cell between the two. Therefore, we have chosen the $\alpha$-cristobalite system for performing the defect calculations.}.

\subsubsection{The disordered interface}
\label{subsubsec:methods-interface-disorder}

For the more realistic disordered interface, we employed the canonical CRN model for network glasses, first used in conjunction with MC techniques to simulate Si~\cite{www,Si-crn} and Ge~\cite{www}, and adapted later for the Si--SiO$_2$ system~\cite{Si-SiO2-crn,Si-SiO2-crn2,Si-SiO2-crn3}. The MC simulations make use of a Keating-like force field~\cite{keating} and the single bond-switch trial move proposed by Wooten, Winter and Weaire (WWW)~\cite{www}. The model was parametrized by fitting the force field to DFT total energy calculations of the entire Si--SiO$_2$ interface supercell; to the best of our knowledge, this is the first fully {\em ab initio} application of the CRN-MC method to this system.

\begin{figure}
\begin{center}
{\small{\begin{picture}(5040.00,3780.00)%
    \gdef\gplbacktext{}%
    \gdef\gplfronttext{}%
    \gplgaddtomacro\gplbacktext{%
      \csname LTb\endcsname%
      \put(814,704){\makebox(0,0)[r]{\strut{}-15}}%
      \put(814,1106){\makebox(0,0)[r]{\strut{}-10}}%
      \put(814,1507){\makebox(0,0)[r]{\strut{}-5}}%
      \put(814,1909){\makebox(0,0)[r]{\strut{} 0}}%
      \put(814,2310){\makebox(0,0)[r]{\strut{} 5}}%
      \put(814,2712){\makebox(0,0)[r]{\strut{} 10}}%
      \put(814,3113){\makebox(0,0)[r]{\strut{} 15}}%
      \put(814,3515){\makebox(0,0)[r]{\strut{} 20}}%
      \put(946,484){\makebox(0,0){\strut{} 0}}%
      \put(1685,484){\makebox(0,0){\strut{} 20}}%
      \put(2425,484){\makebox(0,0){\strut{} 40}}%
      \put(3164,484){\makebox(0,0){\strut{} 60}}%
      \put(3904,484){\makebox(0,0){\strut{} 80}}%
      \put(4643,484){\makebox(0,0){\strut{} 100}}%
      \put(176,2109){\rotatebox{-270}{\makebox(0,0){\strut{}Energy (eV)}}}%
      \put(2794,154){\makebox(0,0){\strut{}MC moves}}%
    }%
    \gplgaddtomacro\gplfronttext{%
      \csname LTb\endcsname%
      \put(3656,1317){\makebox(0,0)[r]{\strut{}DFT}}%
      \csname LTb\endcsname%
      \put(3656,1097){\makebox(0,0)[r]{\strut{}CRN (initial)}}%
      \csname LTb\endcsname%
      \put(3656,877){\makebox(0,0)[r]{\strut{}CRN (fitted)}}%
    }%
    \gplgaddtomacro\gplbacktext{%
    }%
    \gplgaddtomacro\gplfronttext{%
    }%
    \gplbacktext
    \put(0,0){\includegraphics{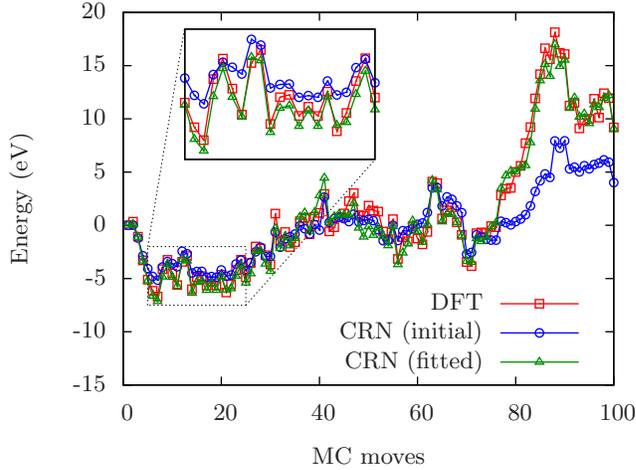}}%
    \gplfronttext
  \end{picture}}}
\caption{Fitting of the CRN model parameters to DFT results for 100 snapshots in a MC run. The total energy for each snapshot is shown for DFT, the initial parametrization of the force field used in the CRN model~\cite{Si-SiO2-crn2}, and the new parametrization obtained by our fitting procedure. In each case the energy is given with respect to that of the initial snapshot.}
\label{fig:int_fit}
\end{center}
\end{figure}

The fitting was performed using an energy-matching procedure~\cite{paul-fit}. The snapshots were obtained from an initial MC run at high temperature ($k_\mathrm{B} T=0.5$~eV) using a parametrization from a previous study~\cite{Si-SiO2-crn2}. $N_s=100$ snapshots were taken from the entire run and used to fit the free parameters of the CRN potential by minimizing the normalized root mean square error $S^\mathrm{fit}$ in total energy differences between all pairs of snapshots for the CRN force field with respect to DFT:
\begin{equation}
S^\mathrm{fit}=\frac{\sqrt{\sum_{i,j}^{N_s} \left ( \left ( E^\mathrm{CRN}_i-E^\mathrm{CRN}_j \right ) - \left ( E^\mathrm{DFT}_i-E^\mathrm{DFT}_j \right ) \right )^2}}{\sqrt{\sum_{i,j}^{N_s} \left ( E^\mathrm{DFT}_i-E^\mathrm{DFT}_j \right )^2}},
\end{equation}
where $E^\mathrm{CRN}_i$ is the total energy of the system for snapshot $i$ given by the CRN force field, and $E^\mathrm{DFT}_i$ is the same quantity given by DFT. Fig.~\ref{fig:int_fit} shows the result of this procedure. The values of the total energy from the initial MC run do not agree well with the DFT results; this is especially clear for the last 24 snapshots. However, there is a significant improvement after our reparametrization: 74\% of the snapshots are in better agreement with DFT after the fitting procedure, and the overall error $S^\mathrm{fit}$ decreases from 0.53 to 0.13. Our new values for the parameters of the CRN potential are given in Table~\ref{table:int_parameters}. We note that the fitting is robust, i.e., similar values are recovered by only fitting to the first half of the data set.

\begin{table}
\caption{List of parameters for the Si--SiO$_2$ CRN model, comparing values from a previous study and those calculated by our fitting procedure. Listed are the equilibrium bond lengths ($b_0$) and angles ($\theta_0$), and their weights ($k_b$ and $k_\theta$, respectively); the force field is defined (following Ref.~\cite{Si-SiO2-crn2}) as $E^\mathrm{CRN} \left ( \mathcal{B}, \left \{ \mathbf{r}_i \right \} \right ) = \sum_i^{N_b} k_b^{\left ( i \right )} \left ( b_i - b_0^{\left ( i \right )} \right )^2 + \sum_i^{N_\theta} k_\theta^{\left ( i \right )} \left ( \cos \theta_i - \cos \theta_0^{\left ( i \right )} \right )^2$, where $\mathcal{B}$ is the bonding topology of the network, $\left \{ \mathbf{r}_i \right \}$ are the atomic positions, $N_b$ is the total number of bonds, $N_\theta$ is the total number of angles between bonded atoms, and the superscript $(i)$ indicates the type of bond/angle, as listed in the table. For each trial $\mathcal{B}$ in a CRN-MC run, $E^\mathrm{CRN}$ is minimized with respect to $\left \{ \mathbf{r}_i \right \}$.}
\begin{center}
{\footnotesize
\lineup
\begin{tabular}{@{}lccccccc}
\br
& \multicolumn{2}{c}{$b_0$ (\AA)} && \multicolumn{4}{c}{$\theta_0$ ($^{\circ}$)} \\ \noalign{\smallskip}\cline{2-3}\cline{5-8}\noalign{\smallskip}
                         & Si--Si & Si--O && Si--Si--Si & Si--Si--O & Si--O--Si & O--Si--O \\
\noalign{\smallskip}\hline\noalign{\smallskip}
Ref.~\cite{Si-SiO2-crn2} & 2.35   & 1.60  && 109.5      & 109.5     & 180.0     & 109.5    \\
This study               & 2.33   & 1.58  && 109.5      & 127.1     & 146.4     & 117.0    \\
\noalign{\smallskip}\hline
\hline\noalign{\smallskip}
& \multicolumn{2}{c}{$k_b$ (eV/\AA$^2$)} && \multicolumn{4}{c}{$k_\theta$ (eV)} \\ \noalign{\smallskip}\cline{2-3}\cline{5-8}\noalign{\smallskip}
                         & Si--Si & Si--O && Si--Si--Si & Si--Si--O & Si--O--Si & O--Si--O \\
\noalign{\smallskip}\hline\noalign{\smallskip}
Ref.~\cite{Si-SiO2-crn2} & 4.54   & 13.50 && 1.79       & 1.97      & 0.38      & 2.16     \\
This study               & 4.56   & 14.12 && 0.83       & 2.01      & 2.97      & 2.92     \\
\br
\end{tabular}
\label{table:int_parameters}
}
\end{center}
\end{table}

Our parametrization of the CRN potential was used to generate the final disordered interface structure. The crystalline interface with $\alpha$-cristobalite (Sec.~\ref{subsubsec:methods-interface-crystal}) was used as our initial configuration for an MC run in which the bonding topologies of the entire SiO$_2$ layer and the first two Si monolayers at the interface were allowed to evolve by bond-switching, while the remaining monolayers of Si were fixed in their bulk bonding topology. The system was first equilibrated at high temperature ($k_\mathrm{B} T=0.5$~eV), and then slowly annealed to a lower temperature ($k_\mathrm{B} T=0.1$~eV); this resulted in the system finding a low-energy local minimum with a disordered interface and with the oxide in its amorphous state, as desired. The equilibration was performed for 4000 accepted moves, and the anneal for 1000 accepted moves. Finally, the system was structurally relaxed using DFT to a tight convergence tolerance of $5 \times 10^{-3}$~eV/\AA\ for the maximum ionic force; this resulted in only small adjustments to the ionic positions, and no change in the bonding network. The width of the final disordered SiO$_2$ layer was $\sim$23.4~\AA.

\begin{figure}
\begin{center}
{\small{\begin{picture}(5040.00,3528.00)%
    \gdef\gplbacktext{}%
    \gdef\gplfronttext{}%
    \gplgaddtomacro\gplbacktext{%
      \csname LTb\endcsname%
      \put(946,704){\makebox(0,0)[r]{\strut{} 0.5}}%
      \put(946,1344){\makebox(0,0)[r]{\strut{} 1}}%
      \put(946,1984){\makebox(0,0)[r]{\strut{} 1.5}}%
      \put(946,2623){\makebox(0,0)[r]{\strut{} 2}}%
      \put(946,3263){\makebox(0,0)[r]{\strut{} 2.5}}%
      \put(1078,484){\makebox(0,0){\strut{}-10}}%
      \put(1969,484){\makebox(0,0){\strut{}-5}}%
      \put(2861,484){\makebox(0,0){\strut{} 0}}%
      \put(3752,484){\makebox(0,0){\strut{} 5}}%
      \put(4643,484){\makebox(0,0){\strut{} 10}}%
      \put(176,1983){\rotatebox{-270}{\makebox(0,0){\strut{}Volume (\AA$^3$)}}}%
      \put(2860,154){\makebox(0,0){\strut{}$z$ (\AA)}}%
    }%
    \gplgaddtomacro\gplfronttext{%
      \csname LTb\endcsname%
      \put(1738,3090){\makebox(0,0)[r]{\strut{}Si$^{4+}$}}%
      \csname LTb\endcsname%
      \put(1738,2870){\makebox(0,0)[r]{\strut{}Si$^{3+}$}}%
      \csname LTb\endcsname%
      \put(1738,2650){\makebox(0,0)[r]{\strut{}Si$^{2+}$}}%
      \csname LTb\endcsname%
      \put(1738,2430){\makebox(0,0)[r]{\strut{}Si$^{1+}$}}%
      \csname LTb\endcsname%
      \put(1738,2210){\makebox(0,0)[r]{\strut{}Si$^{0{\color{white} +}}$}}%
    }%
    \gplbacktext
    \put(0,0){\includegraphics{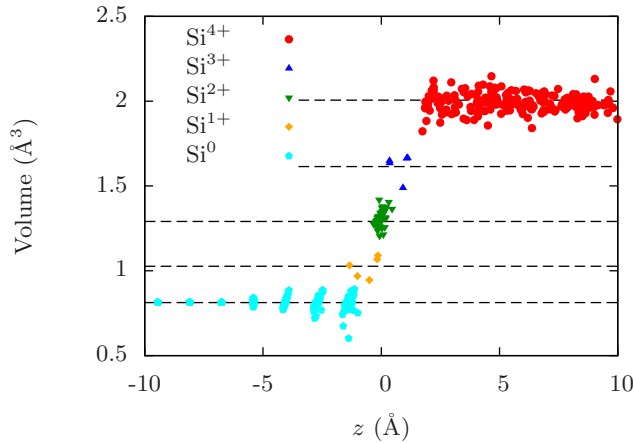}}%
    \gplfronttext
  \end{picture}}}
\caption{Tetrahedral volume of the Si ions in different oxidation states plotted perpendicular to the interface plane. The dashed lines indicate the ideal volume for each oxidation state.}
\label{fig:int_res-ox}
\end{center}
\end{figure}

Although the crystalline configuration used to start the MC simulation only included a single suboxide state ($\mathrm{Si}^{2+}$), the disorder introduced at the interface by the WWW bond-switching mechanism resulted in all the suboxide states being present by the end of the simulation, and none of the O bridges remaining. Fig.~\ref{fig:int_res-ox} shows the distribution and volume of the different oxidation states. The volume is calculated from the tetrahedron formed by the four neighbours of the ion (we use either the middle of the bond for a Si--Si bond, or the O position for a Si--O bond). We also show the `ideal' volume for each state, calculated from the DFT equilibrium bond lengths, giving an indication of the local strain for each ion.

Our model is in good agreement with several key characteristics of the interface that have been established experimentally: the oxide is in its amorphous state~\cite{Si-SiO2-exp-rev}, all oxidation states are present at the interface~\cite{Si-SiO2-sub,Si-SiO2-sub2,Si-SiO2-sub4,Si-SiO2-sub5}, and all atoms are fully coordinated (experimental measurements suggest that there is a very low concentration of over- or under-coordinated atoms, less than one in $10^4$ in the interfacial layers~\cite{Si-SiO2-dangling,Si-SiO2-dangling2}). The thickness of our model interface can be seen from Fig.~\ref{fig:int_res-ox} to be $\sim$5~\AA, close to the lower end of the range of experimental estimates (6--10~\AA\ for the $\mathrm{Si}^{1+}$ and $\mathrm{Si}^{2+}$ states, with the $\mathrm{Si}^{3+}$ state extending further into the oxide~\cite{Si-SiO2-sub,Si-SiO2-sub2,Si-SiO2-sub4,Si-SiO2-sub5}).

\subsection{Arsenic defect calculations}
\label{subsec:methods-As}

Using the crystalline or disordered interfaces discussed above, the energetics of As substitutional defects at Si sites close to and at the interface were calculated using DFT. The technical details for the DFT calculations are the same as those given in Sec.~\ref{subsubsec:methods-interface-crystal}; additionally, for As, we employed a RRKJ norm-conserving pseudopotential that includes the 3d semi-core states. We performed simulations of the system containing a single As impurity; as described in the next section, many different lattice sites were simulated (including all the suboxide states of Si). The segregation energy $E_s$ for a particular site is defined as
\begin{equation} \label{eq:E_s}
E_s = E^\mathrm{def} - E^\mathrm{def,ref},
\end{equation}
where $E^\mathrm{def}$ is the total energy of the supercell with the dopant placed at the site of interest, and $E^\mathrm{def,ref}$ is the same quantity for the dopant placed at a reference site, in this case taken to be at the centre of the Si layer (i.e., furthest away from the interface)~\footnote{We note that the defect formation energy is not well-defined by the Zhang-Northrup~\cite{zhang} approach commonly used in bulk, due to the ambiguity in defining the host chemical potential at inequivalent lattice sites~\cite{nano-Al}; however, this is not a problem for defining the segregation energy, as the number of atoms of each species is constant between the two configurations.}. Only neutral defects are considered; unless otherwise stated, no further relaxation is performed after the impurity is introduced into the system. We do, however, consider the relaxation of the As substitutional defect close to the interface for a smaller sample of lattice sites (Secs.~\ref{subsec:results-rel} and~\ref{subsec:results-strain}).

The calculation of point defect properties with DFT using the supercell approach has some well-known limitations~\cite{Lany2008}, particularly the slow convergence with system size of various quantities of interest, such as the defect formation energy~\cite{Puska1998,Probert2003,Corsetti2011}. However, we expect the segregation energy, being in effect a difference between the formation energy of the substitutional defect at two different lattice sites, to be less sensitive to system size than the formation energy itself.

Although in this study we only consider neutral defects, it is also interesting to note that the definition of the segregation energy given by Eq.~\ref{eq:E_s} in principle remains unchanged when considering the segregation of charged defects, under the assumption that the system is in equilibrium with an electron reservoir provided by the bulk crystal, and, hence, that a global electronic chemical potential can be defined irrespectively of the position of the dopant. This would therefore preclude the need for determining the electronic chemical potential~\cite{Lany2008,Corsetti2011}.

Notwithstanding the limitations, our calculations of the substitutional As defect in a 256-atom supercell of pure bulk Si (similar to the number of atoms in the Si layer in our interface supercell) agree well with experimental results: the ionization energy (calculated as the position of the stable charge transition level $E \left ( {1+} / 0 \right )$ with respect to the conduction band edge) is 47~meV, compared with a value of 49~meV from experiment~\cite{As_Si-ionization}, and the relaxed As--Si bond length is 2.43~\AA, compared with values of $2.41 \pm 0.02$~\AA~\cite{As_Si-bond} and 2.43~\AA~\cite{As_Si-bond2} from experiment. The technical details of the calculations are the same as those given in Sec.~\ref{subsubsec:methods-interface-crystal}, except for the relaxation procedure: the supercell lattice vectors were held constant, and a convergence tolerance of $10^{-3}$~eV/\AA\ for the maximum ionic force was used. These results, therefore, give us confidence in using DFT to investigate As dopants at the Si--SiO$_2$ interface.

\section{Results and discussion}
\label{sec:results}

\subsection{The crystalline interface}
\label{subsec:results-crystal}

\begin{figure}
\begin{center}
\hspace{20pt} \includegraphics[width=0.5\textwidth]{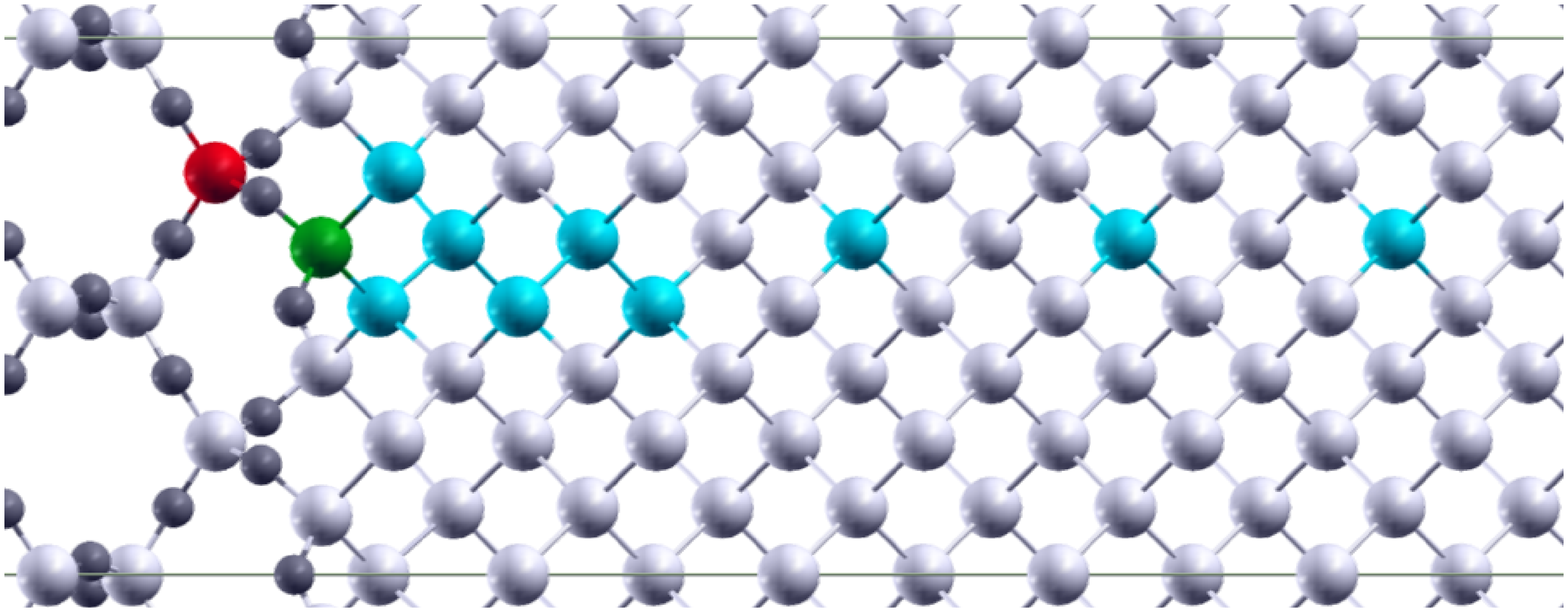}
{\small{\begin{picture}(5040.00,3528.00)%
    \gdef\gplbacktext{}%
    \gdef\gplfronttext{}%
    \gplgaddtomacro\gplbacktext{%
      \csname LTb\endcsname%
      \put(946,704){\makebox(0,0)[r]{\strut{}-0.2}}%
      \put(946,1131){\makebox(0,0)[r]{\strut{}0}}%
      \put(946,1557){\makebox(0,0)[r]{\strut{}0.2}}%
      \put(946,1984){\makebox(0,0)[r]{\strut{}4}}%
      \put(946,2410){\makebox(0,0)[r]{\strut{}8}}%
      \put(946,2837){\makebox(0,0)[r]{\strut{}12}}%
      \put(946,3263){\makebox(0,0)[r]{\strut{}14}}%
      \put(1078,484){\makebox(0,0){\strut{}-5}}%
      \put(1672,484){\makebox(0,0){\strut{} 0}}%
      \put(2266,484){\makebox(0,0){\strut{} 5}}%
      \put(2861,484){\makebox(0,0){\strut{} 10}}%
      \put(3455,484){\makebox(0,0){\strut{} 15}}%
      \put(4049,484){\makebox(0,0){\strut{} 20}}%
      \put(4643,484){\makebox(0,0){\strut{} 25}}%
      \put(176,1983){\rotatebox{-270}{\makebox(0,0){\strut{}$E_s$ (eV)}}}%
      \put(2860,154){\makebox(0,0){\strut{}$z$ (\AA)}}%
    }%
    \gplgaddtomacro\gplfronttext{%
      \csname LTb\endcsname%
      \put(3656,3090){\makebox(0,0)[r]{\strut{}Si$^{4+}$}}%
      \csname LTb\endcsname%
      \put(3656,2870){\makebox(0,0)[r]{\strut{}Si$^{2+}$}}%
      \csname LTb\endcsname%
      \put(3656,2650){\makebox(0,0)[r]{\strut{}Si$^{0{\color{white} +}}$}}%
      \csname LTb\endcsname%
      \put(4465,1283){\makebox(0,0){\strut{}$E_a$}}%
      \put(4465,998){\makebox(0,0){\strut{}$\Delta E$}}%
    }%
    \gplbacktext
    \put(0,0){\includegraphics{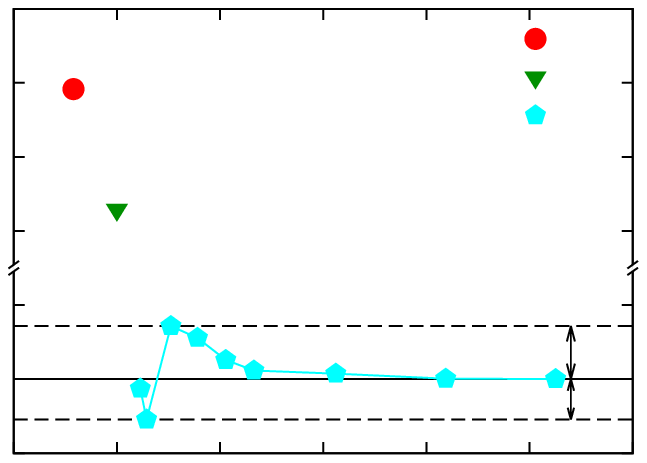}}%
    \gplfronttext
  \end{picture}}}
\caption{Segregation energy of the As dopant at the crystalline interface with respect to its bulk position (bottom panel). The zero of position is taken to be that of the $\mathrm{Si}^{2+}$ site. $E_a$ is the activation energy and $\Delta E$ the maximum segregation energy. Part of the supercell is shown in the top panel. Si atoms are white, and O atoms black; the substitutional sites for the As atom are coloured, with the specific colour denoting its oxidation state. We note that the picture does not show the full supercell.}
\label{fig:seg-ideal}
\end{center}
\end{figure}
 
Due to the ordered atomic arrangement, there are only a small number of inequivalent defect sites, generally one per monolayer; the only important exception is the Si monolayer immediately beneath the first O-bonded layer, which features two inequivalent sites (Fig.~\ref{fig:seg-ideal}, top panel). Consequently, the relaxation of the interface results in a difference of 0.30~\AA\ in the $z$ direction (perpendicular to the interface plane) in the position of these two sites. The local strain (based on the deviation from the ideal volume) is also different, both in sign and magnitude: for the site closer to the interface it is slightly positive, with a volume (6.52~$\mathrm{\AA}^3$) almost identical to the equilibrium one found in bulk (6.50~$\mathrm{\AA}^3$), while for the site further from the interface the strain is negative, with a significantly smaller volume (6.15~$\mathrm{\AA}^3$)~\footnote{Volumes are calculated using the tetrahedron formed by the positions of the four neighbouring ions.}.

Fig.~\ref{fig:seg-ideal} (bottom panel) shows the As segregation energy as a function of the distance to the interface, including the $\mathrm{Si}^{2+}$ site at the interface and the first $\mathrm{Si}^{4+}$ site in SiO$_2$. Direct bonding of the dopant atom to O is highly unfavourable, with an energy penalty of 5.17~eV for the doubly O-coordinated defect site and 11.67~eV for the fully O-coordinated one. This is in agreement with previous theoretical studies~\cite{seg-comp,seg-comp2,seg-comp3,seg-comp4,seg-comp5,seg-comp6}, as well as the experimental evidence that segregated atoms are found on the Si side of the interface.

\begin{figure}
\begin{center}
\begin{overpic}[height=\textwidth,angle=90]{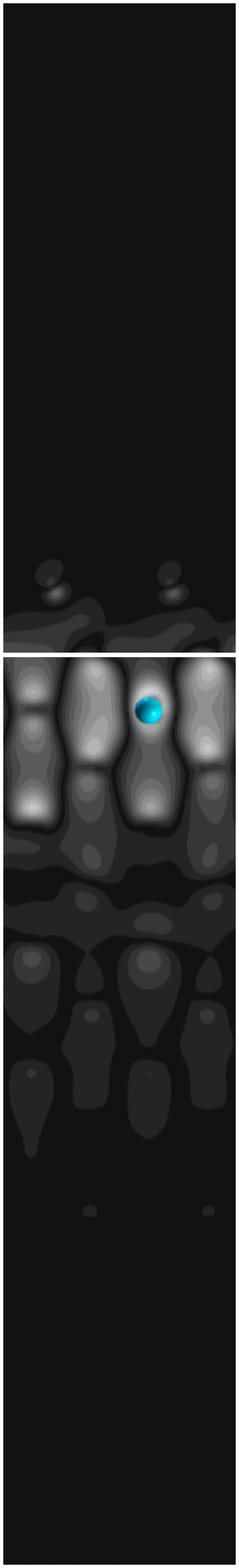}
\put(10,7){\color{yellow} SiO$_2$}
\put(90,7){\color{yellow} Si}
\end{overpic}
{\small{\begin{picture}(5040.00,3528.00)%
    \gdef\gplbacktext{}%
    \gdef\gplfronttext{}%
    \gplgaddtomacro\gplbacktext{%
      \csname LTb\endcsname%
      \put(682,704){\makebox(0,0)[r]{\strut{} 0}}%
      \put(682,1344){\makebox(0,0)[r]{\strut{} 1}}%
      \put(682,1984){\makebox(0,0)[r]{\strut{} 2}}%
      \put(682,2623){\makebox(0,0)[r]{\strut{} 3}}%
      \put(682,3263){\makebox(0,0)[r]{\strut{} 4}}%
      \put(814,484){\makebox(0,0){\strut{} 0}}%
      \put(1404,484){\makebox(0,0){\strut{} 5}}%
      \put(1994,484){\makebox(0,0){\strut{} 10}}%
      \put(2583,484){\makebox(0,0){\strut{} 15}}%
      \put(3173,484){\makebox(0,0){\strut{} 20}}%
      \put(3763,484){\makebox(0,0){\strut{} 25}}%
      \put(3895,704){\makebox(0,0)[l]{\strut{} 30}}%
      \put(3895,1344){\makebox(0,0)[l]{\strut{} 35}}%
      \put(3895,1984){\makebox(0,0)[l]{\strut{} 40}}%
      \put(3895,2623){\makebox(0,0)[l]{\strut{} 45}}%
      \put(3895,3263){\makebox(0,0)[l]{\strut{} 50}}%
      \put(176,1983){\rotatebox{-270}{\makebox(0,0){\strut{}$\Delta z$ (\AA)}}}%
      \put(4532,1983){\rotatebox{-270}{\makebox(0,0){\strut{}$Q$ ($\mathrm{\AA}^2$)}}}%
      \put(2288,154){\makebox(0,0){\strut{}$z$ (\AA)}}%
    }%
    \gplgaddtomacro\gplfronttext{%
      \csname LTb\endcsname%
      \put(2776,3090){\makebox(0,0)[r]{\strut{}$\Delta z$}}%
      \csname LTb\endcsname%
      \put(2776,2870){\makebox(0,0)[r]{\strut{}$Q$}}%
    }%
    \gplbacktext
    \put(0,0){\includegraphics{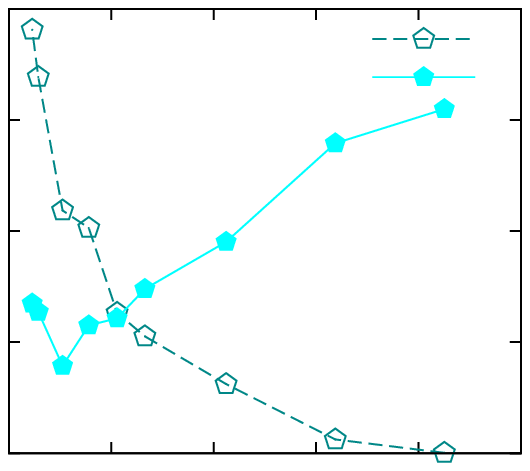}}%
    \gplfronttext
  \end{picture}}}
\caption{Spread $Q$ of the dopant donor level charge density distribution and displacement of its centre from the As position $\Delta z$ at different sites in the crystalline interface system (bottom panel). The values of $\Delta z$ are positive, as the charge distribution is shifted away from the SiO$_2$ region ($z<0$). An example charge density distribution contour map for a defect site close to the interface in the z-y plane is shown in the top panel; the vertical white line indicates the position of the interface. A square root scale is used, with lighter regions indicating a higher density.}
\label{fig:seg-ideal2}
\end{center}
\end{figure}

Restricting ourselves to the fully Si-coordinated defect sites, we see two separate regimes: firstly, a long-ranged steady increase in $E_s$ as the As ion approaches the interface from bulk Si; this reaches its maximum value in the second $\mathrm{Si}^0$ monolayer from the interface, with a small energy penalty of 0.14~eV. Secondly, a drop in $E_s$ at the first $\mathrm{Si}^0$ monolayer for both sites previously described; these sites, therefore, can drive dopant segregation to the interface. We can consider the energy barrier at the second monolayer as a migration or activation energy for the segregation process. We note, however, that the energy gain from segregation is modest: 0.11~eV for the site with a large negative strain, and only 0.03~eV for that with a small positive strain.

All the defect sites belonging to the first regime have a negligible strain, being almost perfectly bulk-like in their local bonding environment both in terms of bond lengths and angles. We therefore assume that the increase in $E_s$ is a confinement effect due to the SiO$_2$, affecting the long-range decay of the defect wavefunction. This effect is analogous to that shown in DFT studies of semiconducting nanocrystals~\cite{quant-conf-doped2} and nanowires~\cite{E_f-conf-nanowire}, in which the defect formation energy for a substitutional dopant atom increases as the system size decreases. Furthermore, we can quantify the confinement by calculating the quadratic spread of the charge density associated with the donor level eigenstate; this is shown in Fig.~\ref{fig:seg-ideal2} for all the $\mathrm{Si}^0$ defect sites. The spread decreases monotonically (up to the second $\mathrm{Si}^0$ monolayer) as the As ion approaches the interface; the top panel of Fig.~\ref{fig:seg-ideal2} shows that this is due to the charge distribution being confined by the interface, resulting in a long-range decay into the bulk Si region and an abrupt decay into the SiO$_2$. This asymmetric decay also results in the centre of the charge density being shifted from the position of the defect site and away from the interface. The calculated magnitude of the shift in the $z$ direction increases as the As ion approaches the interface (bottom panel of Fig.~\ref{fig:seg-ideal2}).

\begin{figure}
\begin{center}
\includegraphics[width=\textwidth]{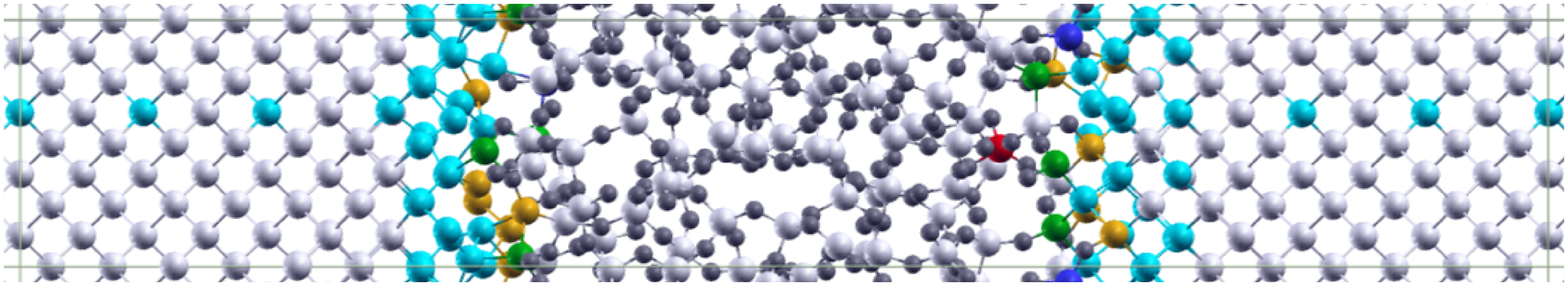} \\
{\small{\begin{picture}(5040.00,3528.00)%
    \gdef\gplbacktext{}%
    \gdef\gplfronttext{}%
    \gplgaddtomacro\gplbacktext{%
      \csname LTb\endcsname%
      \put(946,704){\makebox(0,0)[r]{\strut{}-0.2}}%
      \put(946,1344){\makebox(0,0)[r]{\strut{} 0}}%
      \put(946,1984){\makebox(0,0)[r]{\strut{} 0.2}}%
      \put(946,2623){\makebox(0,0)[r]{\strut{} 0.4}}%
      \put(946,3263){\makebox(0,0)[r]{\strut{} 0.6}}%
      \put(1078,484){\makebox(0,0){\strut{} 10}}%
      \put(1791,484){\makebox(0,0){\strut{} 15}}%
      \put(2504,484){\makebox(0,0){\strut{} 20}}%
      \put(3217,484){\makebox(0,0){\strut{} 25}}%
      \put(3930,484){\makebox(0,0){\strut{} 30}}%
      \put(4643,484){\makebox(0,0){\strut{} 35}}%
      \put(176,1983){\rotatebox{-270}{\makebox(0,0){\strut{}$E_s$ (eV)}}}%
      \put(2860,154){\makebox(0,0){\strut{}$z$ (\AA)}}%
    }%
    \gplgaddtomacro\gplfronttext{%
    }%
    \gplgaddtomacro\gplbacktext{%
      \csname LTb\endcsname%
      \put(2846,1996){\makebox(0,0)[r]{\strut{} {\scriptsize 0}}}%
      \put(2846,2504){\makebox(0,0)[r]{\strut{} {\scriptsize 10}}}%
      \put(2846,3011){\makebox(0,0)[r]{\strut{} {\scriptsize 20}}}%
      \put(2886,1877){\makebox(0,0){\strut{}{\scriptsize -0.2}}}%
      \put(3235,1877){\makebox(0,0){\strut{} {\scriptsize 0}}}%
      \put(3585,1877){\makebox(0,0){\strut{} {\scriptsize 0.2}}}%
      \put(3934,1877){\makebox(0,0){\strut{} {\scriptsize 0.4}}}%
      \put(4283,1877){\makebox(0,0){\strut{} {\scriptsize 0.6}}}%
      \put(2578,2503){\rotatebox{-270}{\makebox(0,0){\strut{}{\scriptsize $N_s$}}}}%
      \put(3584,1666){\makebox(0,0){\strut{}{\scriptsize $E_s$ (eV)}}}%
    }%
    \gplgaddtomacro\gplfronttext{%
    }%
    \gplbacktext
    \put(0,0){\includegraphics{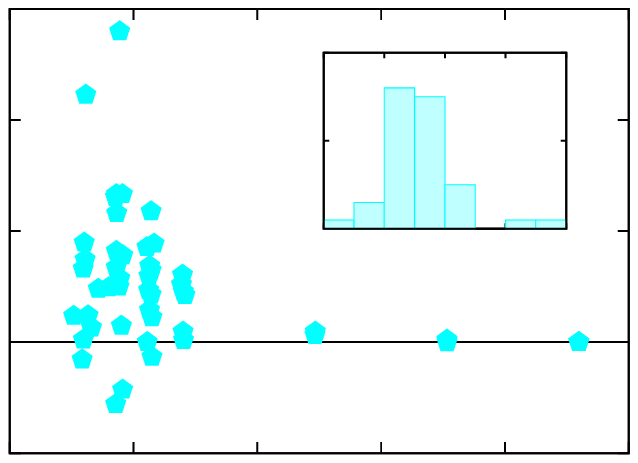}}%
    \gplfronttext
  \end{picture}}}
\caption{Segregation energy of the As dopant in $\mathrm{Si}^0$ sites of the disordered interface system (bottom panel). The zero of position is taken to be the centre of the oxide layer. The inset shows a histogram of the number of sites $N_s$ at each segregation energy for the first four Si monolayers below the interface. The supercell used for the calculations is shown in the top panel, following the colour scheme of Fig.~\ref{fig:seg-ideal}.}
\label{fig:seg-amorph3}
\end{center}
\end{figure}

The crystalline interface, therefore, clearly illustrates the large-scale confinement of the defect wavefunction at the interface, which has a relatively small effect on $E_s$ (the ionization energy, instead, is expected to be greatly influenced, as discussed in previous studies~\cite{nanowire-ionization,arsenic-ionization}). This effect depends only on the distance of the dopant to the interface, instead of on the detailed local ionic configuration around the defect site. It is this local bonding, however, that has the greatest effect on $E_s$, as demonstrated by the abrupt change between the first two $\mathrm{Si}^0$ monolayers.

\subsection{The disordered interface}
\label{subsec:results-disorder}

In the disordered interface system with the amorphous oxide, there are a large number of inequivalent sites for all suboxide states. This enables detailed investigation of the effect of defect volume and local strain on the As segregation energy; we perform calculations on 67 sites in total. The disorder effectively masks the small energy barrier to segregation discussed in the previous section: Fig.~\ref{fig:seg-amorph3} shows that there is no clear dependence of $E_s$ on $z$ close to the interface. In the first few monolayers of Si, we find four sites with a negative value of $E_s$ (all Si$^0$).

\begin{figure}
\begin{center}
{\small{\begin{picture}(7200.00,3528.00)%
    \gdef\gplbacktext{}%
    \gdef\gplfronttext{}%
    \gplgaddtomacro\gplbacktext{%
      \csname LTb\endcsname%
      \put(876,888){\makebox(0,0)[r]{\strut{}0}}%
      \put(876,1619){\makebox(0,0)[r]{\strut{}4}}%
      \put(876,2351){\makebox(0,0)[r]{\strut{}8}}%
      \put(876,3082){\makebox(0,0)[r]{\strut{}12}}%
      \put(1008,485){\makebox(0,0){\strut{}1.2}}%
      \put(1735,485){\makebox(0,0){\strut{}1.4}}%
      \put(2462,485){\makebox(0,0){\strut{}1.6}}%
      \put(3189,485){\makebox(0,0){\strut{}1.8}}%
      \put(3916,485){\makebox(0,0){\strut{}2}}%
      \put(370,1985){\rotatebox{-270}{\makebox(0,0){\strut{}$E_s$ (eV)}}}%
      \put(2462,155){\makebox(0,0){\strut{}$V^{1/3}$ (\AA)}}%
    }%
    \gplgaddtomacro\gplfronttext{%
    }%
    \gplgaddtomacro\gplbacktext{%
      \csname LTb\endcsname%
      \put(4072,888){\makebox(0,0)[r]{\strut{}}}%
      \put(4072,1619){\makebox(0,0)[r]{\strut{}}}%
      \put(4072,2351){\makebox(0,0)[r]{\strut{}}}%
      \put(4072,3082){\makebox(0,0)[r]{\strut{}}}%
      \put(4204,485){\makebox(0,0){\strut{}1.2}}%
      \put(4931,485){\makebox(0,0){\strut{}1.4}}%
      \put(5658,485){\makebox(0,0){\strut{}1.6}}%
      \put(6385,485){\makebox(0,0){\strut{}1.8}}%
      \put(7112,485){\makebox(0,0){\strut{}2}}%
      \put(5658,155){\makebox(0,0){\strut{}$V^{1/3}$ (\AA)}}%
    }%
    \gplgaddtomacro\gplfronttext{%
      \csname LTb\endcsname%
      \put(6125,3092){\makebox(0,0)[r]{\strut{}Si$^{4+}$}}%
      \csname LTb\endcsname%
      \put(6125,2872){\makebox(0,0)[r]{\strut{}Si$^{3+}$}}%
      \csname LTb\endcsname%
      \put(6125,2652){\makebox(0,0)[r]{\strut{}Si$^{2+}$}}%
      \csname LTb\endcsname%
      \put(6125,2432){\makebox(0,0)[r]{\strut{}Si$^{1+}$}}%
      \csname LTb\endcsname%
      \put(6125,2212){\makebox(0,0)[r]{\strut{}Si$^{0{\color{white} +}}$}}%
    }%
    \gplbacktext
    \put(0,0){\includegraphics{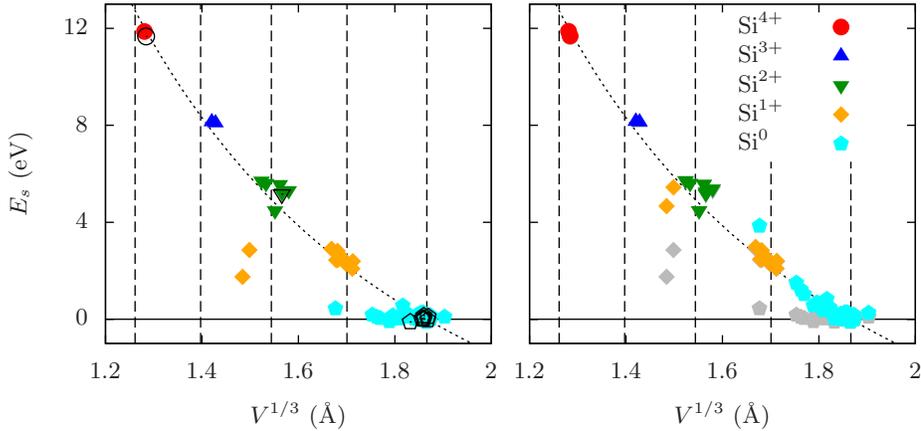}}%
    \gplfronttext
  \end{picture}}}
\caption{Segregation energy of the As dopant at the disordered interface with respect to its bulk position, as a function of the cube root of the volume of the defect site $V$. The vertical dashed lines indicate the ideal volume for each oxidation state. Filled coloured symbols indicate sites in the disordered system, and empty black symbols ones in the crystalline system. The fine dashed line is a function of the form $V^{-2/3}$ fitted to the data points for the crystalline system only. In the left panel, the unadjusted data points are shown; in the right panel, the same data points are shown with the addition of a strain energy term (as explained in the text), with the unadjusted values now appearing in light grey.}
\label{fig:seg-uncorrcorr}
\end{center}
\end{figure}

In order to understand the large spread of values of $E_s$ in the disordered system, we consider local quantities: the volume of the defect site $V$, and the local volumetric strain at that site $e_V^\mathrm{loc} = \left ( V-V_0^{(i)} \right )/V_0^{(i)}$ (where $V_0^{(i)}$ is the ideal volume for oxidation state $i$). Fig.~\ref{fig:seg-uncorrcorr} (left) shows $E_s$ for all the simulated defect sites as a function of the cube root of the defect volume, both for the crystalline and disordered systems. Based on a simple `particle in a box' confinement model for the defect charge density, we fit a function of the form $V^{-2/3}$ to the data points obtained from the crystalline system only. The majority of data points for the disordered system are in good agreement with this curve; in particular, the $\mathrm{Si}^{1+}$ sites follow it closely for a small range of volumes. The ideal volume for each oxidation state is also indicated, showing that most defect sites have little strain. However, there are some outlying data points, mainly for $\mathrm{Si}^0$ and $\mathrm{Si}^{1+}$; it is interesting to note that these all correspond to highly strained configurations, and have noticeably lower energies than those predicted by the fitted curve. A possible explanation, which we consider here, is that the segregation energy is lowered by the presence of local strain at the defect site, since the system effectively gains energy by removing the Si ion from this strained site to an unstrained bulk site. Using a simple harmonic approximation for the strain energy of the form $B V_0^{(i)} {e_V^\mathrm{loc}}^2/2$, we can perform a second fit to describe the deviation of the outliers from the main curve. Fig.~\ref{fig:seg-uncorrcorr} (right) shows that subtracting the fitted value for the strain energy is quite successful in bringing the outliers back onto the curve while having a negligible effect on the other points; however, the value of the coefficient $B$ obtained from the fitting ($\approx 4000$~GPa) is extremely high compared with realistic bulk moduli, casting doubt on its physical significance.

\begin{table}
\caption{Summary table of results obtained for the lattice relaxation of the As impurity in the interface supercells. The oxidation state and position in the supercell (either in the bulk-like region in the middle of the Si layer or in the interface region) are listed for each defect site considered. Also listed are whether the system is relaxed or not, the bond lengths between the As ion and its four neighbours \{a, b, c, d\}, the local volumetric strain at the defect site, and the segregation energy. Asterisks (*) denote an As--O bond. Note that the segregation energy is always given referenced to the unrelaxed bulk Si$^0$ site in the crystalline system.}
\begin{center}
{\footnotesize
\lineup
\begin{tabular}{@{}ccccccccccc}
\br
&&&& \multicolumn{4}{c}{Bond lengths (\AA)} &&& \\ \noalign{\smallskip}\cline{5-8}\noalign{\smallskip}
System & Ox. & Pos. & Rel. & a & b & c & d & $V$ ($\mathrm{\AA}^3$) & $e_V^\mathrm{loc}$ & $E_s$ (eV) \\
\noalign{\smallskip}\hline\noalign{\smallskip}
\multirow{6}{*}{Crystal.}   & \multirow{4}{*}{Si$^0$}    & \multirow{2}{*}{Bulk} & No  & 2.32 & 2.32 & 2.32{\color{white} *} & 2.32{\color{white} *} & 6.43 & $-$0.01                   & {\color{white} $-$}0.00 \\
                            &                            &                       & Yes & 2.42 & 2.42 & 2.42{\color{white} *} & 2.42{\color{white} *} & 7.25 & {\color{white} $-$}0.12 & $-$0.23 \\
                            &                            & \multirow{2}{*}{Int.} & No  & 2.31 & 2.31 & 2.30{\color{white} *} & 2.30{\color{white} *} & 6.15 & $-$0.05                 & $-$0.11 \\
                            &                            &                       & Yes & 2.39 & 2.38 & 2.38{\color{white} *} & 2.38{\color{white} *} & 6.81 & {\color{white} $-$}0.05 & $-$0.39 \\ \noalign{\smallskip}\cline{2-11}\noalign{\smallskip}
                            & \multirow{2}{*}{Si$^{2+}$} & \multirow{2}{*}{Int.} & No  & 2.35 & 2.31 & 1.65*                 & 1.63*                 & 3.84 & {\color{white} $-$}0.04 & {\color{white} $-$}5.17 \\
                            &                            &                       & Yes & 2.46 & 2.41 & 1.87*                 & 1.85*                 & 4.68 & {\color{white} $-$}0.27 & {\color{white} $-$}3.51 \\
\noalign{\smallskip}\hline\noalign{\smallskip}
\multirow{6}{*}{Disorder.}  & \multirow{4}{*}{Si$^0$}    & \multirow{2}{*}{Bulk} & No  & 2.32 & 2.32 & 2.32{\color{white} *} & 2.32{\color{white} *} & 6.43 & $-$0.01                 & $-$0.01 \\
                            &                            &                       & Yes & 2.42 & 2.42 & 2.42{\color{white} *} & 2.41{\color{white} *} & 7.24 & {\color{white} $-$}0.11 & $-$0.24 \\
                            &                            & \multirow{2}{*}{Int.} & No  & 2.49 & 2.34 & 2.32{\color{white} *} & 2.31{\color{white} *} & 6.49 & $-$0.00                 & $-$0.12 \\
                            &                            &                       & Yes & 2.75 & 2.38 & 2.38{\color{white} *} & 2.37{\color{white} *} & 7.37 & {\color{white} $-$}0.13 & $-$0.50 \\ \noalign{\smallskip}\cline{2-11}\noalign{\smallskip}
                            & \multirow{2}{*}{Si$^{1+}$} & \multirow{2}{*}{Int.} & No  & 2.66 & 2.47 & 2.38{\color{white} *} & 1.66*                 & 3.27 & $-$0.34                 & {\color{white} $-$}1.74 \\
                            &                            &                       & Yes & 2.82 & 2.43 & 2.39{\color{white} *} & 1.94*                 & 3.68 & $-$0.25                 & {\color{white} $-$}0.75 \\
\br
\end{tabular}
\label{table:seg_summary}
}
\end{center}
\end{table}

Clearly, both the form of the fitted equations and the definition of the input variables (local volume and strain) are extremely simplified, and discard a large amount of information about the local configuration. The final model employs only two free parameters in total (the coefficients for the $V^{-2/3}$ and strain energy terms), making the ratio of the number of data points to the number of fitting parameters very high. However, the data points for each oxidation state are quite separate from each other, generally clustering in a small range of $E_s$. If we consider the main contribution to $E_s$ to come simply from an energy penalty due to the Si--O bond being replaced by an As--O bond, we would expect a linear relationship between $E_s$ and oxidation. The $V^{-2/3}$ arguably presents a better fit to the data (this can be seen in Fig.~\ref{fig:seg-uncorrcorr}, as the equilibrium volumes of the various oxidation states are approximately equally spaced), while also accounting for the variation of $E_s$ within a single state. Nevertheless, the model should be considered as semi-qualitative. This is shown by a simple test, in which we manually increase the volume of one of the Si$^0$ defect sites at the interface (allowing the rest of the system to relax), and calculate the change in segregation energy: we find $\left. \partial E_s/\partial V \right |_{V_0} = -0.62$~eV/\AA$^3$ (where $V_0$ is the original site volume), compared with $-0.45$~eV/\AA$^3$ obtained from the model. Although the correct trend is reproduced, the error is substantial.

\subsection{Lattice relaxation}
\label{subsec:results-rel}

So far, we have only considered unrelaxed defect configurations (i.e., although both the crystalline and disordered interfaces are fully relaxed using DFT, no further relaxation is performed once the As ion is introduced). However, the As substitutional defect in bulk Si features a small outwards relaxation (as described in Sec.~\ref{subsec:methods-As}), that might differ in magnitude for sites at the interface due to the additional strain field. Therefore, we perform structural relaxation calculations for six defect supercells in total, three each for the crystalline and disordered interfaces, corresponding to the impurity at a bulk-like site at the centre of the Si layer, at the Si$^0$ site at the interface with the greatest energy gain from segregation, and at a partially oxidized site at the interface (Si$^{2+}$ for the crystalline system and Si$^{1+}$ for the disordered one). Ionic forces were converged to less than $10^{-2}$~eV/\AA; the relaxed configurations and energies are given in Table~\ref{table:seg_summary}.

\begin{figure}
\begin{center}
{\small{\begin{picture}(5040.00,3528.00)%
    \gdef\gplbacktext{}%
    \gdef\gplfronttext{}%
    \gplgaddtomacro\gplbacktext{%
      \csname LTb\endcsname%
      \put(682,1024){\makebox(0,0)[r]{\strut{} 0}}%
      \put(682,1664){\makebox(0,0)[r]{\strut{} 2}}%
      \put(682,2303){\makebox(0,0)[r]{\strut{} 4}}%
      \put(682,2943){\makebox(0,0)[r]{\strut{} 6}}%
      \put(814,484){\makebox(0,0){\strut{} 1.4}}%
      \put(2090,484){\makebox(0,0){\strut{} 1.6}}%
      \put(3367,484){\makebox(0,0){\strut{} 1.8}}%
      \put(4643,484){\makebox(0,0){\strut{} 2}}%
      \put(176,1983){\rotatebox{-270}{\makebox(0,0){\strut{}$E_s$ (eV)}}}%
      \put(2728,154){\makebox(0,0){\strut{}$V^{1/3}$ (\AA)}}%
    }%
    \gplgaddtomacro\gplfronttext{%
      \csname LTb\endcsname%
      \put(3656,3090){\makebox(0,0)[r]{\strut{}Si$^{2+}$}}%
      \csname LTb\endcsname%
      \put(3656,2870){\makebox(0,0)[r]{\strut{}Si$^{1+}$}}%
      \csname LTb\endcsname%
      \put(3656,2650){\makebox(0,0)[r]{\strut{}Si$^{0{\color{white} +}}$}}%
    }%
    \gplbacktext
    \put(0,0){\includegraphics{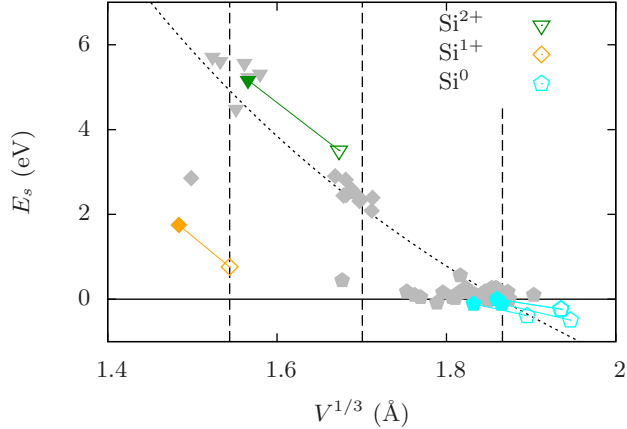}}%
    \gplfronttext
  \end{picture}}}
\caption{Effect of lattice relaxation on the segregation energy. Data points for both the crystalline and disordered systems are shown. Filled coloured symbols indicate the energy and volume before relaxation, and empty coloured symbols after relaxation. The strain energy adjustment is not included.}
\label{fig:seg-rel}
\end{center}
\end{figure}

For all sites, the results show an outwards relaxation and a decrease in the energy of the system on the order of 0.1--1~eV. The energy gain is greater for the Si$^0$ sites at the interfaces that those further in bulk, resulting in an increase in the magnitude of the segregation energy from 0.11~eV to 0.16~eV in the crystalline system, and from 0.11~eV to 0.26~eV in the disordered one. This suggests that lattice relaxation will increase the number of Si$^0$ sites favouring segregation, as some of the sites close to the interface with small positive values of $E_s$ (see inset Fig.~\ref{fig:seg-amorph3}) will become thermodynamically favourable for segregation ($E_s < 0$) upon relaxation.

The relaxed bond lengths and decrease in total energy for the defect site at the centre of the Si layer are in good agreement with those calculated in the 256-atom BCC supercell of bulk Si. As observed previously for the P substitutional defect at the interface~\cite{seg-comp5}, the optimized structure for As at the interface exhibits one As--Si bond that is longer than the other three. This effect is most pronounced in the disordered system, with a difference of 0.37~\AA\ between the long bond and the short ones; however, the dopant is still active after relaxation, as there is a negligible shift in the position of the donor level in the Kohn-Sham band structure (not shown).

Fig.~\ref{fig:seg-rel} shows the effect of the outwards ionic relaxation on the segregation energy and local volume. The Si$^{1+}$ and Si$^{2+}$ sites in particular appear to follow the slope of the previously-fitted $V^{-2/3}$ curve during the relaxation process. These partially oxidized sites gain significantly more energy from relaxation than the Si$^0$ ones; even for Si$^{1+}$, however, this is not sufficient to make segregation to these sites energetically favourable. It seems reasonable to assume that even after taking lattice relaxation into account the number of oxidized sites favouring segregation will be negligible. 

\subsection{Macroscopic strain}
\label{subsec:results-strain}

Finally, we have investigated the possibility of altering the segregation energy for As at interfacial sites through the application of a macroscopic uniform areal strain $e_A$ parallel to the $z$ direction. This areal strain is defined as $e_A = \left ( A-A_0 \right ) / A_0$, where $A$ is the cross-sectional area of the supercell (i.e., the (001) plane for Si), and $A_0$ is the equilibrium area calculated from bulk Si. We restrict ourselves to Si$^0$ sites, and calculate the change in $E_s$ for variations in $A$ of $\pm 2\%$. It is important to note that the reference energy $E^\mathrm{def,ref}$ from Eq.~\ref{eq:E_s} also changes with $e_A$; this change needs to be calculated separately.

We consider all sites found previously with a negative segregation energy: two for the crystalline system, and four for the disordered system. These calculations are performed without relaxation. Additionally, we choose one site each from the crystalline and disordered systems, and perform full ionic relaxation in the presence of the areal strain.

\begin{table}
\caption{Summary table of results for the application of a small areal strain to the interface supercell with an As impurity. Listed are whether the system is relaxed or not, the local volumetric strain at the impurity site, and the change of segregation energy with areal strain.}
\begin{center}
{\footnotesize
\lineup
\begin{tabular}{@{}ccccc}
\br
System                     & Site               & Rel. & $e_V^\mathrm{loc}$      & $\partial E_s/ \partial e_A$ (eV) \\
\noalign{\smallskip}\hline\noalign{\smallskip}
\multirow{3}{*}{Crystal.}  & a                  & No   & {\color{white} $-$}0.00 & {\color{white} $-$}0.25       \\ \noalign{\smallskip}\cline{2-5}\noalign{\smallskip}
                           & \multirow{2}{*}{b} & No   & $-$0.05                 & {\color{white} $-$}0.26       \\
                           &                    & Yes  & {\color{white} $-$}0.05 & {\color{white} $-$}0.22       \\
\noalign{\smallskip}\hline\noalign{\smallskip}
\multirow{6}{*}{Disorder.} & c                  & No   & $-$0.01                 & $-$0.02                       \\ \noalign{\smallskip}\cline{2-5}\noalign{\smallskip}
                           & d                  & No   & $-$0.03                 & $-$0.04                       \\ \noalign{\smallskip}\cline{2-5}\noalign{\smallskip}
                           & e                  & No   & $-$0.12                 & $-$1.38                       \\ \noalign{\smallskip}\cline{2-5}\noalign{\smallskip}
                           & \multirow{2}{*}{f} & No   & $-$0.00                 & $-$0.05                       \\
                           &                    & Yes  & {\color{white} $-$}0.13 & $-$0.57                       \\
\br
\end{tabular}
\label{table:strain}
}
\end{center}
\end{table}

Table~\ref{table:strain} gives the change in $E_s$ with respect to $e_A$ at $e_A=0$ for all sites. The crystalline system shows a small increase in $E_s$ with tensile strain for both sites, with almost no change due to relaxation. The disordered system, instead, shows a decrease with tensile strain for all sites. Interestingly, sites with a small local strain exhibit a negligible effect from the macroscopic areal strain, while those with a large local strain exhibit an effect that is larger by up to two orders of magnitude. Therefore, the application of tensile strain is predicted to favour the segregation of As to the interface, by increasing the stability of Si$^0$ sites in highly strained local configurations, which are naturally found predominantly in the interfacial region.

\section{Conclusions}
\label{sec:outro}

In summary, we have studied the energetics of neutral substitutional As defects at Si sites close to and at the Si--SiO$_2$ interface. Our simulations on both a crystalline and a realistic disordered interface have revealed quite a rich and complex behaviour for this system, characterized by the interplay of various subtle effects. We draw a number of overall conclusions based on our findings:

\begin{itemize}
\item long-range quantum confinement of the defect charge at the interface (affecting the long tail of the weakly bound donor state) results in only a small barrier to segregation, on the order of 0.1~eV;
\item the local bonding environment of the As impurity has a large effect on the segregation energy, with variations on the order of 1~eV for each oxidation state, and is thus more important than long-range effects that depend instead on the macroscopic position of the impurity with respect to the interface;
\item partially and fully O-coordinated Si sites carry a strong energy penalty that makes them unfavourable for As segregation, even after taking into account the substantial lowering of segregation energies by lattice relaxation for such sites;
\item a small number of fully Si-coordinated sites within the first three monolayers below the interface (corresponding to a density of $\sim$0.02~$\mathrm{\AA}^{-2}$) are found to be energetically favourable with respect to the bulk site, with the addition of lattice relaxation potentially further increasing this number;
\item As dopants remain electrically active after segregation to Si substitutional sites, and, therefore, can be used for nanoscale devices;
\item the application of macroscopic tensile strain is observed to increase segregation to locally strained sites in the disordered interface, thereby allowing for a possible measure of control over the segregation process in experimental devices.
\end{itemize}

\ack

This work was supported by the UK Engineering and Physical Sciences Research Council (EPSRC). The calculations were performed on cx2 (Imperial College London High Performance Computing Service) and HECToR (UK National Supercomputing Service). We thank the UK's HPC Materials Chemistry Consortium (EPSRC Grant EP/F067496/1) and Car-Parrinello Consortium (EPSRC Grant EP/K013564/1) for access to HECToR.

\end{document}